\documentstyle[12pt]{article}
 
\newcommand{\beq}[1]{\begin{equation}\label{#1}}
\newcommand{\eeq}{\end{equation}}
\newcommand{\bear}[1]{\begin{eqnarray}\label{#1}}
\newcommand{\ear}{\end{eqnarray}}
\newcommand{\nn}{\nonumber}

\catcode`\@=11 \@addtoreset{equation}{section}\catcode`\@=12
\newcommand{\be}{\begin{equation}}
\newcommand{\ee}{\end{equation}}
\newcommand{\ba}{\begin{eqnarray}}
\newcommand{\ea}{\end{eqnarray}}

\newcommand{\np}{ {\newpage } }

\newcommand{\N}{ \mbox{\rm I$\!$N} }
\newcommand{\R}{ \mbox{\rm I$\!$R} }
\def\C{\mbox{\rm {I\kern-.520em C}}}

\newcommand{\sign}{ \mbox{\rm sign} }
\newcommand{\e}{ \mbox{\rm e} }
\newcommand{\eps}{ \varepsilon }

\newcommand{\p}{\partial}
\newcommand{\btd}{\bigtriangledown}
\newcommand{\btu}{\bigtriangleup}
\newcommand{\tri}{\Delta}

\newcommand{\sums}{\sum\limits}
\newcommand{\const}{\mathop{\rm const}\nolimits}

\newcommand{\unb}[1]{\underbrace{#1}}
\newcommand{\sh}{\mathop{\rm sh}\nolimits}
\newcommand{\ch}{\mathop{\rm ch}\nolimits}
\newcommand{\im}{{\rm i}}

\begin{document}

\begin{center} \large \bf
Multidimensional Classical and Quantum Cosmology\\
with Intersecting $p$-branes
\end{center}

\vspace{1.03truecm}

\bigskip

\centerline{\bf \large
V. D. Ivashchuk and V. N. Melnikov}

\vspace{0.96truecm}

\centerline{Center for Gravitation and Fundamental Metrology}
\centerline{VNIIMS, 3-1 M. Ulyanovoy Str.}
\centerline{Moscow, 117313, Russia}
\centerline{e-mail: melnikov@fund.phys.msu.su}

\begin{abstract}

Multidimensional cosmological model describing the evolution of $(n+1)$
Einstein spaces in the theory with several scalar fields and forms is
considered. When a (electro-magnetic composite)
$p$-brane Ansatz is adopted the field equations are
reduced to the equations for Toda-like system. The Wheeler--De Witt 
equation is obtained. In the case when $n$ "internal" spaces are Ricci-flat,
one space $M_0$ has a non-zero curvature, and all $p$-branes do not "live"
in $M_0$, the classical and quantum solutions are obtained if certain
orthogonality relations on parameters are imposed. 
Spherically-symmetric solutions 
with intersecting non-extremal $p$-branes are singled out.
A non-orthogonal generalization of intersection rules corresponding 
to (open, closed) Toda lattices 
is obtained. A chain of  bosonic $D \geq 11$ models
(that may be related to hypothetical  higher dimensional
supergravities  and $F$-theories) is suggested.

\end{abstract}

\hspace*{0.950cm} PACS number(s):\ 04.50.+h,\ 98.80.Hw,\ 04.60.Kz
\np

\section{\bf Introduction}
\setcounter{equation}{0}

In this paper we continue our investigations of $p$-brane solutions
(see for example \cite{DKL,St,AIR} and references therein)
based on sigma-model approach   \cite{IM4,IM,IMR}
. (For pure gravitational sector see \cite{RZ,IM0}.)

Here we consider a cosmological case, when all functions 
depend upon one variable (time). The model under consideration
contains several scalar fields and antisymmetric forms and is
governed by action (\ref{2.1i}).

The considered cosmological model contains some stringy cosmological
models (see for example \cite{LMPX}. It may be obtained (at classical
level) from multidimensional cosmological model with perfect fluid
\cite{IM3}-\cite{GIM} as an interesting special case. 

The paper is organized as follows. In Sect. 2 the model with composite 
electro-magnetic p-branes is described. In Sect. 3 the
$\sigma$-model representation (under certain constraints on $p$-branes)
is considered and certain scalar invariants are calculated. In Sect. 4
the Wheeler-DeWitt equation is obtained. Sect. 5 is devoted to exact 
solutions (classical and quantum) for orthogonal case, when one 
factor space is curved. In Subsect. 5.3  spherically symmetric
configurations with non-extremal $p$-branes is singled out.
In Sect. 6 the intersection rules are generalized to non-orthogonal
case and a chain of bosonic $B_D$-models ($D \geq 11$) containing 
"truncated" $D= 11$ supergavity model \cite{CJS} and $D = 12$ model from
\cite{KKP} is suggested. 
(It may be supposed that
these models or their analogues may be connected 
with higher dimensional generalization of $M$- and $F$-theories 
\cite{S,D,HTW,HV}.)

\section{\bf The model}
\setcounter{equation}{0}

Here like in \cite{IM} we consider the model governed by the action
\bear{2.1i}
S =&& \frac{1}{2\kappa^{2}}
\int_{M} d^{D}z \sqrt{|g|} \{ {R}[g] - 2 \Lambda - h_{\alpha\beta}\;
g^{MN} \partial_{M} \varphi^\alpha \partial_{N} \varphi^\beta
\\ \nn
&& - \sum_{a \in \Delta}
\frac{\theta_a}{n_a!} \exp[ 2 \lambda_{a} (\varphi) ] (F^a)^2_g \}
+ S_{GH},
\ear
where $g = g_{MN} dz^{M} \otimes dz^{N}$ is the metric
($M,N =1, \ldots, D$), $\varphi=(\varphi^\alpha)\in \R^l$
is a vector from dilatonic scalar fields,
$(h_{\alpha\beta})$ is a non-degenerate $l\times l$ matrix ($l\in \N$),
$\theta_a = \pm 1$,
\beq{2.2i}
F^a =  dA^a =
\frac{1}{n_a!} F^a_{M_1 \ldots M_{n_a}}
dz^{M_1} \wedge \ldots \wedge dz^{M_{n_a}}
\eeq
is a $n_a$-form ($n_a \geq 1$) on a $D$-dimensional manifold $M$,
$\Lambda$ is cosmological constant and $\lambda_{a}$ is a $1$-form
on $\R^l$: $\lambda_{a} (\varphi) =\lambda_{a \alpha} \varphi^\alpha$,
$a \in \Delta$, $\alpha=1,\ldots,l$. In (\ref{2.1i}) we denote
$|g| = |\det (g_{MN})|$,
\beq{2.3i}
(F^a)^2_g =
F^a_{M_1 \ldots M_{n_a}} F^a_{N_1 \ldots N_{n_a}}
g^{M_1 N_1} \ldots g^{M_{n_a} N_{n_a}},
\eeq
$a \in \Delta$, where $\Delta$ is some finite set, and $S_{\rm GH}$ is the
standard Gibbons-Hawking boundary term \cite{GH}. In the models
with one time all $\theta_a =  1$  when the signature of the metric
is $(-1,+1, \ldots, +1)$.

The equations of motion corresponding to  (\ref{2.1i}) have the following
form
\bear{2.4i}
R_{MN} - \frac{1}{2} g_{MN} R  =   T_{MN} - \Lambda g_{MN},
\\
\label{2.5i}
{\btu}[g] \varphi^\alpha -
\sum_{a \in \Delta} \theta_a  \frac{\lambda^{\alpha}_a}{n_a!}
e^{2 \lambda_{a}(\varphi)} (F^a)^2_g = 0,
\\
\label{2.6i}
\nabla_{M_1}[g] (e^{2 \lambda_{a}(\varphi)}
F^{a, M_1 \ldots M_{n_a}})  =  0,
\ear
$a \in \Delta$; $\alpha=1,\ldots,l$.
In (\ref{2.5i}) $\lambda^{\alpha}_{a} = h^{\alpha \beta}
\lambda_{a \beta }$, where $(h^{\alpha \beta})$
is matrix inverse to $(h_{\alpha \beta})$.
In (\ref{2.4i})
\bear{2.7i}
T_{MN} =   T_{MN}[\varphi,g]
+ \sum_{a\in\Delta} \theta_a  e^{2 \lambda_{a}(\varphi)} T_{MN}[F^a,g],
\ear
where
\bear{2.8i}
T_{MN}[\varphi,g] =
h_{\alpha\beta}\left(\p_{M} \varphi^\alpha \p_{N} \varphi^\beta -
\frac{1}{2} g_{MN} \p_{P} \varphi^\alpha \p^{P} \varphi^\beta\right),
\\
T_{MN}[F^a,g] = \frac{1}{n_{a}!}\left[ - \frac{1}{2} g_{MN} (F^{a})^{2}_{g}
+ n_{a}  F^{a}_{M M_2 \ldots M_{n_a}} F_{N}^{a, M_2 \ldots M_{n_a}}\right].
\label{2.9i}
\ear
In (\ref{2.5i}), (\ref{2.6i}) ${\btu}[g]$ and ${\btd}[g]$
are Laplace-Beltrami and covariant derivative operators respectively
corresponding to  $g$.

Let us consider the manifold
\beq{2.10g}
M = \R  \times M_{0} \times \ldots \times M_{n}
\eeq
with the metric
\beq{2.11g}
g= w \e^{2{\gamma}(u)} du \otimes du +
\sum_{i=0}^{n} \e^{2\phi^i(u)} g^i ,
\eeq
where $w=\pm 1$, $u$ is a distinguished coordinate which, by
convention, will be called ``time";
$g^i  = g^i_{m_{i} n_{i}}(y_i) dy_i^{m_{i}} \otimes dy_i^{n_{i}}$
is a metric on $M_{i}$  satisfying the equation
\beq{2.12g}
R_{m_{i}n_{i}}[g^i ] = \xi_{i} g^i_{m_{i}n_{i}},
\eeq
$m_{i},n_{i}=1,\ldots,d_{i}$; $d_{i} = \dim M_i$, $\xi_i= \const$,
$i=0,\dots,n$; $n\in {\bf N}$. Thus, $(M_i,g^i)$ are Einstein spaces.
The functions $\gamma,\phi^i$: $(u_-,u_+)\to{\bf R}$ are smooth.

Each manifold $M_i$ is assumed to be oriented and connected,
$i = 0,\ldots,n$. Then the volume $d_i$-form
\beq{2.13g}
\tau_i  = \sqrt{|g^i(y_i)|}
\ dy_i^{1} \wedge \ldots \wedge dy_i^{d_i},
\eeq
and the signature parameter
\beq{2.14g}
\eps(i)  = \sign \det (g^i_{m_{i}n_{i}}) = \pm 1
\eeq
are correctly defined for all $i=0,\ldots,n$.

Let
\beq{2.15g}
\Omega_0 = \{ \emptyset, \{ 0 \}, \{ 1 \}, \ldots, \{ n \},
\{ 0, 1 \}, \ldots, \{ 0, 1,  \ldots, n \} \}
\eeq
be a set of all subsets of
\beq{2.25n}
I_0\equiv\{ 0, \ldots, n \}.
\eeq
For any $I = \{ i_1, \ldots, i_k \} \in \Omega_0$, $i_1 < \ldots < i_k$,
we define a form
\beq{2.17i}
\tau(I) \equiv \tau_{i_1}  \wedge \ldots \wedge \tau_{i_k},
\eeq
of rank
\beq{2.19i}
d(I) \equiv  \sum_{i \in I} d_i = d_{i_1} + \ldots + d_{i_k},
\eeq
and a corresponding $p$-brane submanifold
\beq{2.18i}
M_{I} \equiv M_{i_1}  \times  \ldots \times M_{i_k},
\eeq
where $p=d(I)-1$ (${\rm dim M_{I}} = d(I)$).
We also define $\eps$-symbol 
\beq{2.19e}
\eps(I) \equiv  \eps(i_1) \ldots \eps(i_k).
\eeq
For $I = \emptyset$ we put  $\tau(\emptyset) = \eps(\emptyset) = 1$,
 $d(\emptyset) = 0$.

For fields of forms we adopt the following "composite electro-magnetic"
Ansatz
\beq{2.27n}
F^a=\sum_{I\in\Omega_{a,e}}F^{(a,e,I)}+\sum_{J\in\Omega_{a,m}}F^{(a,m,J)},
\eeq
where
\bear{2.28n}
F^{(a,e,I)}=d\Phi^{(a,e,I)}\wedge\tau(I), \\ \label{2.29n}
F^{(a,m,J)}=\e^{-2\lambda_a(\varphi)}
*\left(d\Phi^{(a,m,J)}\wedge\tau(J)\right),
\ear
$a\in\tri$, $I\in\Omega_{a,e}$, $J\in\Omega_{a,m}$ and
\beq{2.29nn}
\Omega_{a,e},\Omega_{a,m}\subset \Omega_0.
\eeq
(For empty $\Omega_{a,v}=\emptyset$, $v=e,m$, we put $\sums_\emptyset=0$ in
(\ref{2.27n})). In (\ref{2.29n}) $*=*[g]$ is the Hodge operator on $(M,g)$.

For the potentials in (\ref{2.28n}), (\ref{2.29n}) we put
\beq{2.28nn}
\Phi^s=\Phi^s(u),
\eeq
$s\in S$, where
\beq{6.39i}
S=S_e\sqcup S_m,  \qquad
S_v\equiv \coprod_{a\in\tri}\{a\}\times\{v\}\times\Omega_{a,v},
\eeq
$v=e,m$.

For dilatonic scalar fields we put
\beq{2.30n}
\varphi^\alpha=\varphi^\alpha(u),
\eeq
$\alpha=1,\dots,l$.

>From  (\ref{2.28n})  and (\ref{2.29n}) we obtain 
the relations between dimensions of $p$-brane 
worldsheets and ranks of forms
\bear{2.d1}
d(I) = n_a - 1,  \quad I \in \Omega_{a,e},
\\ \label{2.d2}
d(J) = D - n_a - 1,  \quad J \in \Omega_{a,m},
\ear
in electric and magnetic cases respectively.

\section{\bf $\sigma$-model representation}
\setcounter{equation}{0}

{\bf Restrictions on $\Omega_{a,v}$.} Let
\beq{4.12i}
w_1\equiv\{i \mid i\in\{1,\dots,n\},\ d_i=1\}.
\eeq
The set $w_1$ describes all $1$-dimensional manifolds among $M_i$ $(i\ge0)$.
We impose the following restrictions on the sets $\Omega_{a,v}$
(\ref{2.29nn}):
\beq{4.31in}
W_{ij}(\Omega_{a,v})=\emptyset,
\eeq
$a\in\tri$; $v=e,m$; $i,j\in w_1$, $i<j$ and
\beq{6.17in}
W_j^{(1)}(\Omega_{a,m},\Omega_{a,e})=\emptyset,
\eeq
$a\in\tri$; $j\in w_1$. Here
\bear{4.23i}
W_{ij}(\Omega_*)\equiv
\{(I,J)|I,J\in\Omega_*,\ I=\{i\}\sqcup(I\cap J),\
J=\{j\}\sqcup(I\cap J)\},
\ear
$i,j\in w_1$, $i\ne j$, $\Omega_* \subset \Omega_0$ and
\beq{6.17i}
W_j^{(1)}(\Omega_{a,m},\Omega_{a,e}) \equiv
\{(I,J)\in\Omega_{a,m}\times\Omega_{a,e}|\bar I=\{j\}\sqcup J\},
\eeq
$j\in w_1$. In (\ref{6.17i})
\beq{5.28i}
\bar I\equiv I_0 \setminus I
\eeq
is "dual" set. (The restrictions (\ref{4.31in}) and (\ref{6.17in}) are
trivially satisfied when $n_1\le1$ and $n_1=0$ respectively, where
$n_1=|w_1|$ is the number of $1$-dimensional manifolds among $M_i$).

It follows from \cite{IM} (see Proposition 2 in \cite{IM}) that the
equations of motion (\ref{2.4i})--(\ref{2.6i}) and the Bianchi
identities
\beq{2.b}
dF^s=0, \quad s\in S
\eeq
for the field configuration (\ref{2.11g}), (\ref{2.27n})--(\ref{2.28nn}),
(\ref{2.30n}) with the restrictions (\ref{4.31in}), (\ref{6.17in})
imposed are equivalent to equations of motion for $\sigma$-model
with the action
\beq{2.25gn}
S_{\sigma} = \frac{\mu}2
\int du {\cal N} \biggl\{G_{ij}\dot\phi^i\dot\phi^j
+h_{\alpha\beta}\dot\varphi^{\alpha}\dot\varphi^{\beta} 
+\sum_{s\in S}\eps_s\exp[-2U^s(\phi,\varphi)](\dot\Phi^s)^2
-2{\cal N}^{-2}V(\phi)\biggr\},
\eeq
where $\dot x\equiv dx/du$,
\beq{2.27gn}
V = {V}(\phi) = -w\Lambda\e^{2\gamma_0(\phi)}+
\frac w2\sum_{i =0}^{n} \xi_i d_i \e^{-2 \phi^i + 2 {\gamma_0}(\phi)}
\eeq
is the potential with
\beq{2.24gn}
\gamma_0(\phi)
\equiv\sum_{i=0}^nd_i\phi^i,  \label{2.32g}
\eeq
and
\beq{2.24gn1}
{\cal N}=\exp(\gamma_0-\gamma)>0
\eeq
is the lapse function,
\bear{2.u}
 U^s = U^s(\phi,\varphi)= -\chi_s\lambda_{a_s}(\varphi) +
\sum_{i\in I_s}d_i\phi^i, \\ \label{2.e}
\eps_s=(-\eps[g])^{(1-\chi_s)/2}\eps(I_s)\theta_{a_s}
\ear
for $s=(a_s,v_s,I_s)\in S$, $\eps[g]= \sign \det (g_{MN})$,
\bear{2.x1}
\chi_s=+1, \quad v_s=e; \\ \label{2.x2}
\chi_s=-1, \quad v_s=m,
\ear
and
\beq{2.c}
G_{ij}=d_i\delta_{ij}-d_id_j
\eeq
are components of the "pure cosmological" minisupermetric, $i,j=0,\dots,n$
\cite{IMZ}.

In the electric case $(F^{(a,m,I)}=0)$ for finite internal space
volumes $V_i$ the action (\ref{2.25gn}) coincides with the
action (\ref{2.1i}) if 
$\mu=-w/\kappa_0^2$, $\kappa^{2} = \kappa^{2}_0 V_0 \ldots V_n$.

Action (\ref{2.25gn}) may be also written in the form
\beq{2.31n}
S_\sigma=\frac\mu2\int du{\cal N}\left\{
{\cal G}_{\hat A\hat B}(X)\dot X^{\hat A}\dot X^{\hat B}-
2{\cal N}^{-2}V(X)\right\},
\eeq
where $X = (X^{\hat A})=(\phi^i,\varphi^\alpha,\Phi^s)\in{\bf 
R}^{N}$, and minisupermetric \beq{2.31nn} {\cal G}={\cal G}_{\hat 
A\hat B}(X)dX^{\hat A}\otimes dX^{\hat B} \eeq on minisuperspace 
\beq{2.31m}
{\cal M}={\bf R}^{N}, \quad   N = n+1+l+|S|
\eeq
($|S|$ is the number of elements in $S$) is defined by the relation
\beq{2.33n}
({\cal G}_{\hat A\hat B}(X))=\left(\begin{array}{ccc}
G_{ij}&0&0\\[5pt]
0&h_{\alpha\beta}&0\\[5pt]
0&0&\eps_s\e^{-2U^s(X)}\delta_{ss'}
\end{array}\right).
\eeq

\subsection{\bf Scalar products}

The minisuperspace metric (\ref{2.31nn}) may be also written in the form
\beq{2.34n}
{\cal G}=\bar G+\sum_{s\in S}\eps_s\e^{-2U^s(x)}d\Phi^s\otimes d\Phi^s,
\eeq
where $x=(x^A)=(\phi^i,\varphi^\alpha)$,
\bear{2.35n}
\bar G=\bar G_{AB}dx^A\otimes dx^B=G_{ij}d\phi^i\otimes d\phi^j+
h_{\alpha\beta}d\varphi^\alpha\otimes d\varphi^\beta, \\ \label{2.36n}
(\bar G_{AB})=\left(\begin{array}{cc}
G_{ij}&0\\
0&h_{\alpha\beta}
\end{array}\right),
\ear
$U^s(x)=U_A^sx^A$ is defined in  (\ref{2.u}) and
\beq{2.38n}
(U_A^s)=(d_i\delta_{iI_s},-\chi_s\lambda_{a_s\alpha}).
\eeq
Here
\beq{2.39n}
\delta_{iI}\equiv\sum_{j\in I}\delta_{ij}=\begin{array}{ll}
1,&i\in I\\
0,&i\notin I
\end{array}
\eeq
is an indicator of $i$ belonging to $I$. The potential (\ref{2.27gn})
reads
\beq{2.40n}
V=(-w\Lambda)\e^{2U^\Lambda(x)}+\sum_{j=0}^n\frac w2\xi_jd_j
\e^{2U^j(x)},
\eeq
where
\bear{2.41n}
U^j(x)=U_A^jx^A=-\phi^j+\gamma_0(\phi), \\ \label{2.42n}
U^\Lambda(x)=U_A^\Lambda x^A=\gamma_0(\phi), \\ \label{2.43n}
(U_A^j)=(-\delta_i^j+d_i,0), \\ \label{2.44n}
(U_A^\Lambda)=(d_i,0).
\ear

The integrability of the Lagrange system (\ref{2.31n}) crucially depends
upon the scalar products of co-vectors $U^\Lambda$, $U^j$, $U^s$
corresponding to $\bar G$:
\beq{2.45n}
(U,U')=\bar G^{AB}U_AU'_B,
\eeq
where
\beq{2.46n}
(\bar G^{AB})=\left(\begin{array}{cc}
G^{ij}&0\\
0&h^{\alpha\beta}
\end{array}\right)
\eeq
is matrix inverse to (\ref{2.36n}). Here (as in \cite{IMZ})
\beq{2.47n}
G^{ij}=\frac{\delta^{ij}}{d_i}+\frac1{2-D},
\eeq
$i,j=0,\dots,n$. These
products have the following form
\bear{2.48n}
(U^i,U^j)=\frac{\delta_{ij}}{d_j}-1, \\ \label{2.49n}
(U^i,U^\Lambda)=-1, \\ \label{2.50n}
(U^\Lambda,U^\Lambda)=-\frac{D-1}{D-2}, \\ \label{2.51n}
(U^s,U^{s'})=q(I_s,I_{s'})+\chi_s\chi_{s'}
\lambda_{a_s}\cdot\lambda_{a_{s'}}, \\ \label{2.52n}
(U^s,U^i)=-\delta_{iI_s}, \\ \label{2.53n}
(U^s,U^\Lambda)=\frac{d(I_s)}{2-D},
\ear
where $s=(a_s,v_s,I_s)$, $s'=(a_{s'},v_{s'},I_{s'})\in S$,
\bear{2.54n}
q(I,J)\equiv d(I\cap J)+\frac{d(I)d(J)}{2-D}, \\ \label{2.55n}
\lambda_a\cdot\lambda_b\equiv\lambda_{a\alpha}\lambda_{b\beta}
h^{\alpha\beta}.
\ear
Relations (\ref{2.48n})-(\ref{2.50n})  were found in 
\cite{GIM}  and (\ref{2.51n})   in \cite{IM}.

\section{\bf Wheeler--De Witt equation}
\setcounter{equation}{0}

Here we fix the gauge as follows
\beq{4.1n}
\gamma_0-\gamma=f(X),  \quad  {\cal N} = e^f,
\eeq
where $f$: ${\cal M}\to{\bf R}$ is a smooth function. Then we obtain the
Lagrange system with the Lagrangian
\beq{3.14r}
L_f=\frac\mu2\e^f{\cal G}_{\hat A\hat B}(X)
\dot X^{\hat A}\dot X^{\hat B}-\mu\e^{-f}V
\eeq
and the energy constraint
\beq{3.15r}
E_f=\frac\mu2\e^f{\cal G}_{\hat A\hat B}(X)
\dot X^{\hat A}\dot X^{\hat B}+\mu\e^{-f}V=
0.
\eeq

Using the standard prescriptions of (covariant and conformally
covarint)  quantization (see, for example,
\cite{Mis,Hal,IMZ}) we are led to the Wheeler-DeWitt (WDW) equation
\beq{4.2n}
\hat{H}^f \Psi^f \equiv
\left(-\frac{1}{2\mu}\Delta\left[e^f{\cal G}\right]+
\frac{a}{\mu}R\left[e^f{\cal G}\right]
+e^{-f}\mu V\right)\Psi^f=0,
\eeq
where
\beq{4.3n}
a=a_N=\frac{N-2}{8(N-1)}. 
\eeq
Here $\Psi^f=\Psi^f(X)$ is the so-called "wave function of the universe"
corresponding to the $f$-gauge (\ref{4.1n}), 
and satisfying the relation 
\beq{4.3an}
\Psi^f= e^{bf} \Psi^{f=0}, \quad b = (2-N)/2,
\eeq
($\Delta[{\cal G}_1]$ and
$R[{\cal G}_1]$ denote the Laplace-Beltrami operator and the scalar
curvature corresponding to ${\cal G}_1$).

For the scalar curvature of minisupermetric  (\ref{2.34n})
we get (see (2.29) in \cite{IM})
\beq{4.4n}
R[{\cal G}]=-\sum_{s\in S}(U^s,U^s)-
\sum_{s,s'\in S}(U^s,U^{s'}).
\eeq

For the Laplace operator we obtain
\beq{4.5n}
\tri[{\cal G}]
=\e^{U(x)}\frac\partial{\partial x^A}\left(\bar G^{AB}
\e^{-U(x)}\frac\partial{\partial x^B}\right) 
+\sum_{s\in S}\eps_s\e^{2U^s(x)}
\left(\frac\partial{\partial\Phi^s}\right)^2,
\eeq
where  $U(x)=\sum_{s\in S}U^{s}(x)$.

{\bf Harmonic-time gauge.} The WDW equation (\ref{4.2n}) for $f=0$
\beq{4.7n}
\hat H\Psi\equiv\left(-\frac{1}{2\mu}\Delta[{\cal G}]+
\frac{a}{\mu}R[{\cal G}]+\mu V\right)\Psi=0,
\eeq
may be rewritten, using relations (\ref{4.4n}), (\ref{4.5n}) and
\beq{4.8n}
U^{si}= G^{ij}U_j^s= \delta_{iI_s}-\frac{d(I_s)}{D-2}, \quad
U^{s\alpha}= - \chi_s \lambda_{a_s}^\alpha,
\eeq
as follows
\bear{4.10n}
2\mu\hat H\Psi=\left\{-G^{ij}\frac\partial{\partial\phi^i}
\frac\partial{\partial\phi^j}-h^{\alpha\beta}
\frac\partial{\partial\varphi^\alpha}\frac\partial{\partial\varphi^\beta}-
\sum_{s\in S}\eps_s\e^{2U^s(\phi,\varphi)}
\left(\frac\partial{\partial\Phi^s}\right)^2\right. \\ \nn \left.
+\sum_{s\in S}\left[\sum_{i\in I_s}\frac\partial{\partial\phi^i}-
\frac{d(I_s)}{D-2}\sum_{j=0}^n\frac\partial{\partial\phi^j}-
\chi_s\lambda_{a_s}^\alpha\frac\partial{\partial\varphi^\alpha}\right]+
2aR[{\cal G}]+2\mu^2V\right\}\Psi=0.
\ear
Here $\hat H\equiv\hat H^{f=0}$ and $\Psi\equiv\Psi^{f=0}$.

\section{\bf Exact solutions with one curved factor space and
orthogonal $U^s$}
\setcounter{equation}{0}

Here we put the following restrictions on the parameters of the model
\beq{5.1n}
{\bf (i)}\qquad \Lambda=0,
\eeq
i.e. the cosmological constant is zero,
\beq{5.2n}
{\bf (ii)}\qquad \xi_0\ne0, \quad \xi_1=\dots=\xi_n=0,
\eeq
i.e. one space is curved and others are Ricci-flat,
\beq{5.3n}
{\bf (iii)}\qquad 0\notin I_s, \quad \forall s=(a_s,v_s,I_s)\in S,
\eeq
i.e. all "brane" submanifolds $M_{I_s}$ (see (\ref{2.18i})) do not
contain $M_0$.

We also put the following orthogonality restriction on the vectors $U^s$
\beq{5.4n}
{\bf (iv)} \
(U^s,U^{s'})=d(I_s\cap I_{s'})+\frac{d(I_s)d(I_{s'})}{2-D}+
\chi_s\chi_{s'}\lambda_{a_s\alpha}\lambda_{a_{s'}\beta}h^{\alpha\beta}=0,
\eeq
for $s\ne s'$, $s=(a_s,v_s,I_s)$, $s'=(a_{s'},v_{s'},I_{s'})\in S$ and
\beq{5.5n}
{\bf (v)}\qquad (U^s,U^s)\ne0
\eeq
for all $s\in S$.

>From {\bf (i)}, {\bf (ii)}  we get for the potential (\ref{2.40n})
\beq{5.6n}
V=\frac12w\xi_0d_0\e^{2U^0(x)},
\eeq
where
\beq{5.7n}
(U^0,U^0)=\frac1{d_0}-1<0
\eeq
(see (\ref{2.48n})). 

>From {\bf (iii)} and (\ref{2.52n}) we get
\beq{5.8n}
(U^0,U^{s})=0
\eeq
for all $s\in S$.
Thus, all co-vectors $U^0$, $U^s$ $(s\in S)$ belonging to dual space
$({\bf R}^{n+1+l})^*\simeq{\bf R}^{n+1+l}$ are orthogonal with respect
to the scalar product (\ref{2.45n}).

Let
\beq{5.9n}
\eta_s \equiv \sign(U^s,U^s),
\eeq
and $|S|_\pm$ is the number of $U^s$ with $\eta_s=\pm1$; $|S|_++|S|_-=|S|$.
Let the matrix $(h_{\alpha\beta})$ has the signature $(\unb{-1,\dots,-1}_{l_-},
\unb{+1,\dots,+1}_{l_+})$ $(l_-+l_+=l)$. The matrix $(G_{ij})$ has a
pseudo-Euclidean signature $(-1,\unb{+1,\dots,+1}_n)$ \cite{IMZ}. Then from
(\ref{5.7n})--(\ref{5.8n}) we obtain  $|S|_-\le l_-$, $|S|_+\le n+l_+$
and hence
\beq{5.11n}
|S| \le n+l.
\eeq
Thus we obtain the restriction on the number of orthogonal (in
minisupermetric (\ref{2.46n})) $p$-brane configurations.

\subsection{\bf Quantum solutions}

The truncated minisuperspace metric (\ref{2.36n}) may be diagonalized
by the linear transformation
\beq{5.12n}
z^A=S^A{}_Bx^B, \quad (z^A)=(z^0,z^a,z^s)
\eeq
as follows
\beq{5.13n}
\bar G=\bar G_{AB}dx^A\otimes dx^B=-dz^0\otimes dz^0+
\sum_{s\in S}\eta_sdz^s\otimes dz^s+dz^a\otimes dz^b\eta_{ab},
\eeq
where $a,b=1,\dots,n+l-|S|$; $\eta_{ab} =\eta_{aa} \delta_{ab};
\eta_{aa}= \pm 1$, and
\bear{5.14n}
q_0z^0=U^0(x), \\
\label{5.15n}
q_sz^s=U^s(x),
\ear
with
\bear{5.16n}
q_0\equiv\sqrt{|(U^0,U^0)|}=\sqrt{1-\frac1{d_0}}>0, \\ \label{5.17n}
q_s=\nu_s^{-1}\equiv\sqrt{|(U^s,U^s)|}=
\sqrt{\left|d(I_s)\left(1+\frac{d(I_s)}{2-D}\right)+
\lambda_{a_s}^2\right|}>0,
\ear
$s=(a_s,v_s,I_s)\in S$.

>From (\ref{4.5n}), (\ref{5.12n}), (\ref{5.13n}) and (\ref{5.15n}) we get
\bear{5.18n}
 \tri[{\cal G}]=-\left(\frac\partial{\partial z^0}\right)^2+
\eta^{ab}\frac\partial{\partial z^a}\frac\partial{\partial z^b}+
\sum_{s\in S}\eta_s\e^{q_sz^s}\frac\partial{\partial z^s}
\left(\e^{-q_sz^s}\frac\partial{\partial z^s}\right) \\ \nn
+\sum_{s\in  S} \eps_s 
\e^{2q_sz^s}\left(\frac\partial{\partial\Phi^s}\right)^2.
\ear 

The relation (\ref{4.4n}) in the orthogonal case reads as
\beq{5.19n}
R[{\cal G}]=-2\sum_{s\in S}(U^s,U^s)=
-2\sum_{s\in S}\eta_sq_s^2.
\eeq

We are seeking the solution to WDW equation (\ref{4.7n}) by the method
of the separation of variables, i.e. we put
\beq{5.20n}
\Psi_*(z)=\Psi_0(z^0)\left(\prod_{s\in S}\Psi_s(z^s)\right)
\e^{\im P_s\Phi^s}\e^{\im p_az^a}.
\eeq
It follows from (\ref{5.18n}) that $\Psi_*(z)$ satisfies WDW equation
(\ref{4.7n}) if
\bear{5.21n}
2\hat H_0\Psi_0\equiv\left\{\left(\frac\partial{\partial z^0}\right)^2
+\mu^2w\xi_0d_0\e^{2q_0z^0}\right\}\Psi_0=2{\cal E}_0\Psi_0; \\ \label{5.22n}
2\hat H_s\Psi_s\equiv\left\{-\eta_s\e^{q_sz^s}\frac\partial{\partial z^s}
\left(\e^{-q_sz^s}\frac\partial{\partial z^s}\right)+
\eps_sP_s^2\e^{2q_sz^s}\right\}\Psi_s=2{\cal E}_s\Psi_s,
\ear
$s\in S$, and
\beq{5.23n}
2{\cal E}_0+\eta^{ab}p_ap_b+2\sum_{s\in S}{\cal E}_s+
2aR[{\cal G}]=0,
\eeq
with $a$ and $R[{\cal G}]$ from (\ref{4.3n}) and (\ref{5.19n}) respectively.

Using the relations from Appendix 1 we obtain linearly independent
solutions to (\ref{5.21n}) and (\ref{5.22n}) respectively
\bear{5.24n}
\Psi_0(z^0)=B_{\omega_0({\cal E}_0)}^0
\left(   \sqrt{-w \mu^2 \xi_0d_0}\frac{\e^{q_0z^0}}{q_0}\right), \\ 
\label{5.25n} \Psi_s(z^s)=\e^{q_sz^s/2}B_{\omega_s({\cal E}_s)}^s 
\left(\sqrt{\eta_s\eps_sP_s^2}\frac{\e^{q_sz^s}}{q_s}\right),
\ear
where
\beq{5.27n}
\omega_0({\cal E}_0)=\sqrt{2{\cal E}_0}/q_0, 
\qquad
\omega_s({\cal E}_s)=\sqrt{\frac14- 2\eta_s{\cal E}_s\nu_s^2},
\eeq
$s\in S$ and $B_\omega^0,B_\omega^s=I_\omega,K_\omega$
are the modified Bessel function.

The general solution of the WDW equation (\ref{4.7n}) is a superposition
of the "separated" solutions (\ref{5.20n}):
\beq{5.28}
\Psi(z)=\sum_B\int dpdPd{\cal E} C(p,P,{\cal E},B)
\Psi_*(z|p,P,{\cal E},B),
\eeq
where $p=(p_a)$, $P=(P_s)$, ${\cal E}=({\cal E}_s)$, $B=(B^0,B^s)$,
$B^0,B^s=I,K$; and 
$\Psi_*=\Psi_*(z|p,P,{\cal E},B)$ is given by relation
(\ref{5.20n}), (\ref{5.24n})--(\ref{5.27n}) with ${\cal E}_0$ from
(\ref{5.23n}). Here $C(p,P,{\cal E},B)$ are smooth enough functions.
In non-composite electric case these solutions were considered
recently in \cite{GrIM}.

\subsection{\bf Classical solutions}

{\bf 5.2.1. Toda-like representation.}

Here we will integrate the Lagrange equations corresponding to the
Lagrangian (\ref{3.14r}) with the energy-constraint (\ref{3.15r}) and
hence we will find classical exact solutions when the restrictions
(\ref{5.1n})--(\ref{5.5n}) are imposed. We put $f=0$, i.e. the harmonic
time gauge is considered.

The problem of integrability may be simplified if we integrate the Maxwell
equations (for $s\in S_e$) and Bianchi identities (for $s\in S_m$):
\bear{5.29n}
\frac d{du}\left(\exp(-2U^s)\dot\Phi^s\right)=0
\Longleftrightarrow
\dot\Phi^s=Q_s \exp(2U^s),
\ear
where $Q_s$ are constants, $s=(a_s,v_s,I_s)\in S$.

Let
\bear{5.30n}
&&Q_s\ne0, \quad s\in S_*; \\ \nn
&&Q_s=0, \quad s\in S\setminus S_*,
\ear
where $S_*\subset S$ is a non-empty subset of $S$.

For fixed $Q=(Q_s,s\in S_*)$ the Lagrange equations for the Lagrangian
(\ref{3.14r}) with $f=0$ corresponding to $(x^A)=(\phi^i,\varphi^\alpha)$,
when equations (\ref{5.29n}) are substituted are equivalent to the Lagrange
equations for the Lagrangian
\beq{5.31n}
L_Q=\frac12\bar G_{AB}\dot x^A\dot x^B-V_Q,
\eeq
where
\beq{5.32n}
V_Q=V+\frac12\sum_{s\in S_*}\eps_sQ_s^2\exp[2U^s(x)],
\eeq
$(\bar G_{AB})$ and $V$ are defined in (\ref{2.36n}) and (\ref{2.27gn})
respectively. The zero-energy constraint (\ref{3.15r}) reads
\beq{5.33n}
E_Q=\frac12\bar G_{AB}\dot x^A\dot x^B+V_Q=0.
\eeq

{\bf 5.2.1. Exact solutions for one curved space and orthogonal $U^s$.}

When the conditions {\bf (i)}--{\bf (v)} are satisfied  exact solutions
for Lagrangian (\ref{5.31n}) with the potential (\ref{5.32n}) and $V$ from
(\ref{5.6n}) could readily obtained using the relations from Appendix 2.

The solutions read:
\beq{5.34n}
x^A(u)=-\frac{U^{0A}}{(U^0,U^0)}\ln |f_0(u-u_0)|-
\sum_{s\in S_*}\frac{U^{sA}}{(U^s,U^s)}\ln |f_s(u-u_s)| + c^A u + 
\bar{c}^A,
\eeq
where $u_0$, $u_s$ are constants, $s\in S_*$. Functions $f_0$ and $f_s$
in (\ref{5.34n}) are the following:
\bear{5.35n}
f_0(\tau)=\left|\frac{\xi_0(d_0-1)}{C_0}\right|^{1/2}
\sh(\sqrt{C_0}\tau),  C_0>0,  \xi_0w>0; \\ \label{5.36n}
\left|\frac{\xi_0(d_0-1)}{C_0}\right|^{1/2}
\sin(\sqrt{|C_0|}\tau),  C_0<0,  \xi_0w>0; \\ \label{5.37n}
\left|\frac{\xi_0(d_0-1)}{C_0}\right|^{1/2}
\ch(\sqrt{C_0}\tau),  C_0>0,  \xi_0w<0; \\ \label{5.38n}
\left|\xi_0(d_0-1)\right|^{1/2}
\tau,  C_0=0,  \xi_0w>0,
\ear
and
\bear{5.39n}
f_s(\tau)=\frac{|Q_s|}{\nu_s|C_s|^{1/2}}\sh(\sqrt{C_s}\tau), \;
C_s>0, \; \eta_s\eps_s<0; \\ \label{5.40n}
\frac{|Q_s|}{\nu_s|C_s|^{1/2}}\sin(\sqrt{|C_s|}\tau), \;
C_s<0, \; \eta_s\eps_s<0; \\ \label{5.41n}
\frac{|Q_s|}{\nu_s|C_s|^{1/2}}\ch(\sqrt{C_s}\tau), \;
C_s>0, \; \eta_s\eps_s>0; \\ \label{5.42n}
\frac{|Q^s|}{\nu_s}\tau, \; C_s=0, \; \eta_s\eps_s<0,
\ear
where $C_0$ and $C_s$ are constants. Here we used the relations 
(\ref{5.7n})-(\ref{5.9n}). The contravariant components $U^{0A}= \bar 
G^{AB} U^0_B$ are
\beq{5.43n}
U^{0i}=-\frac{\delta_0^i}{d_0}, \quad U^{0\alpha}=0.
\eeq
Corresponding relations for $U^{sA}$, $s\in S$, were presented in
(\ref{4.8n}).

Using (\ref{5.34n}), (\ref{5.7n}), (\ref{5.17n}), (\ref{4.8n}) and
(\ref{5.43n}) we obtain
\beq{5.44n}
\phi^i=\frac{\delta_0^i}{1-d_0}\ln |f_0|
+\sum_{s\in S_*}\alpha_s^i \ln |f_s| +
c^iu+\bar c^i,
\eeq
where
\beq{5.45n}
\alpha_s^i=-\eta_s\nu_s^2\left(\delta_{iI_s}-\frac{d(I_s)}{D-2}\right),
\eeq
$s\in S$, and
\beq{5.46n}
\varphi^\alpha=\sum_{s\in S_*}\eta_s\nu_s^2\chi_s\lambda_{a_s}^\alpha
\ln |f_s|+c^\alpha u+\bar c^\alpha,
\eeq
$\alpha=1,\dots,l$.

Vectors $c=(c^A)$ and $\bar c=(\bar c^A)$ satisfy the linear constraint
relations (see Appendix 2)
\bear{5.47n}
U^0(c)= U^0_A c^A = -c^0+\sum_{j=0}^nd_jc^j=0, \\ \label{5.48n}
U^0(\bar c)= U^0_A \bar c^A =
-\bar c^0+\sum_{j=0}^nd_j\bar c^j=0, \\ \label{5.49n}
U^s(c)= U^s_A c^A=
\sum_{i\in I_s}d_ic^i-\chi_s\lambda_{a_s\alpha}c^\alpha=0,
\\ \label{5.50n}
U^s(\bar c)=  U^s_A \bar c^A=
\sum_{i\in I_s}d_i\bar c^i-
\chi_s\lambda_{a_s\alpha}\bar c^\alpha=0,
\ear
$s\in S$. The harmonic gauge function (\ref{2.24gn}) reads
\beq{5.51n}
\gamma_0(\phi) = \frac{d_0}{1-d_0}\ln |f_0|+
\sum_{s\in S_*}\frac{d(I_s)}{D-2}\eta_s\nu_s^2\ln |f_s| +
c^0u+\bar c^0.
\eeq

The zero-energy constraint reads (see Appendix 2)
\beq{5.53n}
E=E_0+\sum_{s\in S_*}E_s+ \frac12 \bar G_{AB}c^Ac^B=0,
\eeq
where $C_0=2E_0(U^0,U^0)$, $C_s=2E_s(U^s,U^s)$.
Using the relations (\ref{5.16n}), (\ref{5.17n}), (\ref{5.47n})
we rewrite (\ref{5.53n}) as
\beq{5.55n}
C_0\frac{d_0}{d_0-1}=\sum_{s\in S_*} C_s\nu_s^2\eta_s+
h_{\alpha\beta}c^\alpha c^\beta+\sum_{i=1}^nd_i(c^i)^2+
\frac1{d_0-1}\left(\sum_{i=1}^nd_ic^i\right)^2.
\eeq

>From relation
\beq{5.56n}
\exp 2 U^s=f_s^{-2},
\eeq
following from (\ref{5.34n}), (\ref{5.4n}),  (\ref{5.8n}),
(\ref{5.49n}), (\ref{5.50n}) we get for
electric-type forms (\ref{2.28n})
\beq{5.57n}
F^s=Q_s f_s^{-2}du\wedge\tau(I_s),
\eeq
$s\in S_e$, and for magnetic-type forms (\ref{2.29n})
\beq{5.58n}
F^s=\e^{-2\lambda_a(\varphi)}
*\left[Q_s f_s^{-2} du \wedge\tau(I_s)\right] =
\bar Q_s \tau(\bar I_s),
\eeq
$s\in S_m$, where  $\bar Q_s=Q_s\eps(I_s)\mu(I_s)w$
and $\mu(I) =\pm1$ is defined by the relation
$\mu(I) du \wedge \tau(I_0)=\tau(\bar I)\wedge du\wedge\tau(I)$.
The relation (\ref{5.58n}) may be readily obtained using the formula
(\ref{5.30n}) from \cite{IM} (for $\gamma=\gamma_0$).

Relations for the metric follows from (\ref{5.44n}), (\ref{5.45n})
and (\ref{5.51n})
\bear{5.63n}
g= \biggl(\prod_{s\in 
S_*}[f_s^2(u-u_s)]^{\eta_s d(I_s)\nu_s^2/(D-2)}\biggr) 
\biggl\{[f_0^2(u-u_0)]^{d_0/(1-d_0)}\e^{2c^0u+2\bar c^0}\\ \nn
\times[wdu\otimes du+f_0^2(u-u_0)g^0]+
\sum_{i\ne0}\Bigl(\prod_{s\in S_*\atop I_s\ni i}
[f_s^2(u-u_s)]^{-\eta_s\nu_s^2}\Bigr)\e^{2c^iu+2\bar c^i}g^i\biggr\}.
\ear

Thus we obtained exact solutions for multidimensional cosmology,
describing the evolution of $(n+1)$ spaces $(M_0,g_0),\dots,(M_n,g_n)$,
where $(M_0,g_0)$ is an Einstein space of non-zero curvature, and
$(M_i,g^i)$ are "internal" Ricci-flat spaces, $i=1,\dots,n$; in the
presence of several scalar fields and forms. The solution is presented
by relations  (\ref{5.46n}), (\ref{5.57n})-(\ref{5.63n})
with the functions $f_0$, $f_s$  defined in
(\ref{5.35n})--(\ref{5.42n}) and the relations on the parameters of
solutions $c^A$, $\bar c^A$ $(A=i,\alpha)$, $C_0$, $C_s$ $(s\in S_*)$,
$\nu_s$, imposed in (\ref{5.47n})--(\ref{5.50n}), 
(\ref{5.55n}), (\ref{5.17n}), respectively.

This solution describes a set of charged (by forms) overlapping
$p$-branes ($p_s=d(I_s)-1$, $s\in S_*$) "living" on submanifolds
$M_{I_s}$ (\ref{2.18i}), where the sets $I_s$ do not contain $0$,
i.e. all $p$-branes live in "internal" Ricci-flat spaces.

The solution is valid if the dimensions of $p$-branes and dilatonic
coupling vector satisfy the relations (\ref{5.4n}).
In non-composite case these solutions were considered
recently in \cite{GrIM,BGIM} (electric case) and \cite{BIM}
(electro-magnetic case). For $n = 1$ see also \cite{LPX,LMPX}.

\subsection{\bf Spherically symmetric solutions}

To illustrate the general solution from Subsect. 5.2 let us consider
also the spherically symmetric case
\beq{5.64}
w = 1, \quad M_0 = S^{d_0} \quad g^0 = d \Omega^2_{d_0},
\eeq
where $d \Omega^2_{d_0}$ is canonical metric on unit sphere 
$S^{d_0}$. We also assume that $M_1 = {\bf R}$, 
$g^1 = - dt \otimes dt$ and 
\beq{5.65}
1 \in I_s, \quad \forall s \in S_*,
\eeq
i. e. all p-branes have common time direction $t$. Let
\beq{5.66}
 \eta_s \varepsilon_s = -1, 
\eeq
$s \in S_*$.
For integration constants we put $\bar{c}^A = 0$,
\bear{5.67}
&&c^A = \bar{\mu} \sum_{r \in \bar{S}} \frac{U^{rA}}{(U^r,U^r)}
      - \bar{\mu} \delta^A_1, \\ 
       \label{5.68}
&&C_0 = C_s =  \bar{\mu}^2,
\ear
where $\bar{\mu} > 0$ and  $\bar{S} = \{ 0 \} \cup S$. Here
$A = (i_A, \alpha_A)$ and $A =1$ means $i_A = 1$.
It may be verified that the restrictions
(\ref{5.47n})-(\ref{5.50n}) and (\ref{5.55n}) are satisfied
identically.

We also introduce new radial variable $R = R(u)$ by relations
\beq{5.69}
\exp( - 2\bar{\mu} u) = 1 - \frac{2\mu}{R^{\bar{d}}},  \quad
\mu = \bar{\mu} \bar{d} >0, \quad  \bar{d} = d_0 -1,
\eeq
and put  $u_0 = 0$, $u_s < 0$,
\beq{5.70}
\frac{|Q_s|}{\bar{\mu} \nu_s} \sinh \beta_s =1, \quad
\beta_s \equiv \bar{\mu}| u_s|, \quad s \in S_*.
\eeq

Then,  solutions for the metric and scalar fields
(see (\ref{5.46n}), (\ref{5.63n})) are  the
following                            
\bear{5.72n}
g= 
\Bigl(\prod_{s \in S_*} H_s^{2 \eta_s d(I_s)\nu_s^2/(D-2)} \Bigr)
\biggl\{ \frac{dR \otimes dR}{1 - 2\mu / R^{\bar d}}
+ R^2  d \Omega^2_{d_0}  \\ \nn
- \Bigl(\prod_{s \in S_*} H_s^{-2 \eta_s \nu_s^2} \Bigr)
 \left(1 - \frac{2\mu}{R^{\bar d }} \right)  dt \otimes dt
+ \sum_{i = 2}^{n} \Bigl(\prod_{s\in S_* \atop I_s \ni i}
  H_s^{-2 \eta_s \nu_s^2} \Bigr) g^i  \biggr\}, \\ 
\label{5.73}
\varphi^\alpha=
\sum_{s\in S_*} \eta_s \nu_s^2 \chi_s \lambda_{a_s}^\alpha
\ln H_s,
\ear
where
\beq{5.74}
H_s = 1 + \frac{{\cal P}_s}{R^{\bar{d}}},
\qquad {\cal P}_s \equiv \frac{|Q_s|\bar{d}}{\nu_s} e^{- \beta_s},
\eeq
$s \in S_*$.

The fields of forms are given by  (\ref{2.28n}), (\ref{2.29n})
with
\bear{5.75}
&&\Phi^s = \frac{\nu_s}{H_s^{'} },  \\
\label{5.76}
&&H_s^{'}= \Bigl(1 - \frac{{\cal P}_s^{'}}{H_s R^{\bar{d}}} \Bigr)^{-1} =
1 + \frac{{\cal P}_s^{'}}{ R^{\bar{d}} + {\cal P}_s - {\cal P}_s^{'} }, \\
\label{5.77}
&&{\cal P}_s^{'} \equiv - \frac{Q_s\bar{d}}{\nu_s}.
\ear
$s \in S_*$. It follows from (\ref{5.70}), (\ref{5.74}) and (\ref{5.77})
that
\beq{5.78}
|{\cal P}_s^{'}| = \frac{\mu}{\sinh \beta_s} = {\cal P}_s e^{\beta_s}
= \sqrt{{\cal P}_s ({\cal P}_s +2\mu)},
\eeq
$s \in S_*$.  

The solutions obtained describe non-extremal charged
intersecting $p$-branes and agree with those 
>from Refs. \cite{CT}, \cite{AIV}, \cite{O} 
($d_1 = \ldots = d_n =1$, $\eta_s = + 1$) and \cite{BIM}
($\eta_s = + 1$, non-composite case).

We note that Hawking temperature corresponding to
the solution is (see also \cite{O,BIM})
\beq{5.79}
T_H =
\frac{\bar{d}}{4 \pi (2 \mu)^{1/\bar{d}}}
\prod_{s \in S_*} 
\left(\frac{2 \mu}{2 \mu + {\cal P}_s}\right)^{\eta_s \nu_s^2}.
\eeq
Recall that $\eta_s \nu_s^2 = (U^s,U^s)^{-1}$.

\subsection{\bf WDW equation with fixed charges}

We may consider also another scheme based on zero-energy
constraint relation (\ref{5.33n}). The corresponding WDW
equation in the harmonic gauge reads
\beq{5.80}
\hat{H}_Q \Psi \equiv
\left(-\frac{1}{2\mu}\ {\bar G}^{AB} 
\frac{\partial}{\partial x^A} \frac{\partial}{\partial x^B} 
+\mu V_Q \right) \Psi=0,
\eeq
where potential $V_Q$ is defined in (\ref{5.32n}). This equation
describes quantum cosmology with classical fields of forms
and quantum scale factors and dilatonic fields. Such approach
is equivalent to the  scheme of quantization of multidimensional
perfect fluid considered in \cite{IM3}. 

Eq. (\ref{5.80}) is readily solved in the orthogonal case
(\ref{5.4n}). The basis of solutions is given by the following replacements
in (\ref{5.20n}), (\ref{5.23n}), (\ref{5.25n}) and  (\ref{5.27n}) 
respectively
\bear{5.81}
P_s  \mapsto Q_s,  \\
\label{5.82} 
2aR[{\cal G}] \mapsto 0, \\
\label{5.83} 
\Psi_s(z^s)  \mapsto B_{\omega_s}^s 
\left(\sqrt{\eta_s\eps_s Q_s^2}\frac{\e^{q_sz^s}}{q_s}\right), \\
\label{5.84} 
\omega_s \mapsto \sqrt{- 2\eta_s{\cal E}_s\nu_s^2}.
\ear

We note that recently such solution for the special case
with one internal space ($n = 1$) and non-composite
$p$-branes was considered in \cite{LMMP}.

\section{\bf Intersection rules for integrable configurations with 
orthogonal and non-orthogonal $U^s$}

In this section we analyze in details the orthogonality relation
(\ref{5.4n}). According to 
(\ref{2.d1}), (\ref{2.d2}), (\ref{5.3n}) and (\ref{5.4n}) there
exist some obstacles for the existence of binary configurations,
i. e. solutions with two $p$-branes. 
For these reason some restrictions on the model (\ref{2.1i})
(e. g. on dimension $D$, ranks of forms $n_a$ and scalar products for
dilatonic couplings $\lambda_{a} \cdot \lambda_{b}$) and
the manifold (\ref{2.10g}) (e.g. on dimensions $d_0,d_1, \ldots d_n$) 
should be imposed. For example, there are no
binary configurations when: i) all 
$\lambda_{a} \cdot \lambda_{b}$   are irrational;
ii) $d_0$ is big enough, i. e. the dimension of "internal" 
space (where all $p$-branes live) $D - d_0 -1$ is too small;
iii) dimensions of factor spaces $d_i$ ($i \geq 1$) do not
fit the ranks of forms $n_a$ (see (\ref{2.d1}) and (\ref{2.d2})).

In subsection 6.1 we propose a sufficiency condition
for the existence of "full" spectrum of $p$-brane pairs
(see Propositions 1, 2). This condition is formulated in terms 
of so-called fundamental matrix of the model.

In subsection 6.2  we suggest a chain of $D \geq 11$ models
("beautiful" models). These models  satisfy  conditions
of Proposition 1 for $d_0 =2$. They may be considered as a nice 
polygon for investigating of solutions with intersecting 
$p$-branes and possible generalization to higher dimensional
supergravitational and
super-$p$-brane models ($F_D$- or $M_D$-theories). 

In subsection 6.3  we  consider the generalization of the orthogonality
relation leading to some other integrable Euclidean Toda-like Lagrangians
and find corresponding intersection rules. This 
result may be considered as a "bridge" between
Toda  lattices (open, closed etc.) and intersecting
$p$-branes. It opens a new area for using of Lie algebras
(e. g. affine Lie algebras etc.) in the 
(classical and quantum) models with $p$-branes.

\subsection{\bf Orthogonal $U^s$}

In Sect. 5 we obtained exact cosmological solutions for the model 
(\ref{2.1i}) with $\Lambda = 0$ defined on the manifold (\ref{2.10g}).
In (\ref{2.10g}) $M_0$ is an Einstein space of non-zero curvature and
hence $d_0 = {\rm dim} M_0 > 1$. All $p$-branes "live" in internal
space
\beq{6.1}
\bar{M} = M_{1} \times \ldots \times M_{n}
\eeq
($(M_i,g^i)$ are Ricci-flat). Here we treat the existence of the
solutions with two $p$-branes. Let us consider such
solution with  set of indices $S_* = \{ s_1,s_2 \}$, $s_1 \neq s_2$,
where $s_1 = (a_{s_1}, v_{s_1}, I_{s_1})$, 
$s_1 = (a_{s_1}, v_{s_1}, I_{s_1})$; $a_{s_1}, a_{s_2} \in \Delta$;
 $v_{s_1}, v_{s_2} \in \{e,m \}$;  $I_{s_1}, I_{s_2} 
 \subset \{1, \ldots, n  \}$.

The orthogonality relation  (\ref{5.4n}) defines the 
intersections rule
\beq{6.2}
d(I_{s_1} \cap I_{s_2}) = \Delta(s_1,s_2),
\eeq
where
\beq{6.3}
\Delta(s_1,s_2) \equiv
\frac{d(I_{s_1}) d(I_{s_2})}{D-2} -
\chi_{s_1} \chi_{s_2} \lambda_{a_{s_1}} \cdot \lambda_{a_{s_2}},
\eeq
$s_1 \neq s_2$.
It follows from  (\ref{6.2})   that in orthogonal case
\beq{6.4}
\Delta(s_1,s_2) \in {\bf Z}_{+}  \equiv \{ 0, 1, 2, \ldots \}.
\eeq

The intersection symbol   (\ref{6.3})  is symmetric:
$\Delta(s_1,s_2) = \Delta(s_2,s_1)$. In orthogonal case
it satisfies the obvious restrictions
\bear{6.5}
\Delta(s_1,s_2) \leq {\rm min}(d(I_{s_1}), d(I_{s_2})),
\\ \label{6.6}
d(I_{s_1} \cup I_{s_2}) = d(I_{s_1}) + d(I_{s_2})
- \Delta(s_1,s_2) \leq D - d_0 -1 = {\rm dim} \bar{M}.
\ear
The latter follows from the inclusion 
$M_{I_{s_1} \cup I_{s_2}} \subset \bar{M}$. 

We recall that dimensions $d(I_{s})$ obey the relations
(\ref{2.d1}) and (\ref{2.d2}) that may be written as follows
\beq{6.7}
d(I_s) =  \bar{D} \bar{\chi}_s +   \bar{n}_{a_s} \chi_s,
\eeq
where $\bar{D} = D - 2$,  $\bar{n}_{a} = n_{a} - 1$ and
$\bar{\chi}_s = \frac{1}{2} (1 - \chi_s) = 0,1$ 
($\chi_s = 1, -1$) for $v_s = e,m$ respectively.

Let 
\beq{6.8}
N(a,b) \equiv
\frac{(n_{a} - 1) (n_{b} - 1)}{D-2} -
 \lambda_{a} \cdot \lambda_b,
\eeq
$a,b \in \Delta$. We call $(N(a,b))$  fundamental matrix
of the model (\ref{2.1i}).

>From (\ref{6.3}) and (\ref{6.7}) 
we obtain
\beq{6.9a}
\Delta(s_1,s_2) =  \bar{D} \bar{\chi}_{s_1} \bar{\chi}_{s_2}
+   \bar{n}_{a_{s_1}} \chi_{s_1} \bar{\chi}_{s_2}
+   \bar{n}_{a_{s_2}} \chi_{s_2} \bar{\chi}_{s_1}
+   N(a_{s_1}, a_{s_2}) \chi_{s_1} \chi_{s_2},
\eeq
or, more explicitly,
\bear{6.9}
\Delta(s_1,s_2) =&& N(a_{s_1},a_{s_2}), \quad v_{s_1} = v_{s_2} = e; \\
\label{6.10}
&&\bar{n}_{a_{s_1}}  - N(a_{s_1},a_{s_2}), 
\quad v_{s_1} = e, v_{s_2} = m; \\
\label{6.11}
&&\bar{D} - \bar{n}_{a_{s_1}} - \bar{n}_{a_{s_2}}  + N(a_{s_1},a_{s_2}), 
\quad v_{s_1} = v_{s_2} = m.
\ear

Thus $N(a,b)$ is the dimension of intersection for two electric p-branes
charged by forms $F^a$ and $F^b$ respectively. Fundamental matrix 
defines the intersection rules in electro-magnetic case too.

{\bf Remark 1.} The matrix (\ref{6.8}) is unchanged if 
$\bar{n}_{a}  \mapsto k \bar{n}_{a}$
$\bar{D}  \mapsto k^2 \bar{D}$, $k \in {\bf N}$.
For example, we may consider the chain of models
with $D = 9k^2 + 2$, $n = {\rm rank} F = 3k +1$ originating
>from truncated bosonic part of $D = 11$ supergravity.

{\bf Proposition 1.} Let matrix (\ref{6.8}) satisfies the following
restrictions
\bear
({\bf A}) \quad N(a_{1},a_{2}) \in {\bf Z}_{+}; \ 
N(a_{1},a_{2})  \geq d_0 - 1;   \nn \\
({\bf B})  \quad \quad
N(a_{1},a_{2})  \leq {\rm min}(n_{a_{1}},n_{a_{2}})    - d_0;   \nn \\
({\bf C})  \quad
N(a_{1},a_{2})  \geq n_{a_{1}} + n_{a_{2}}  - D  + d_0 - 1,   \nn
\ear
for any $a,b \in \Delta$ ($d_0 >1$). Then $\Delta(s_1,s_2)$
>from (\ref{6.9a})  satisfies consistency relations 
(\ref{6.4})-(\ref{6.6}) for any $s_1 \neq s_2$ (with 
$d(I_s)$ from (\ref{6.7})).

{\bf Proof.} Let us consider three cases: 
(ee) $v_{s_1} = v_{s_2} = e$;
(em) $v_{s_1} = e, v_{s_2} = m$;
(mm) $v_{s_1} = v_{s_2} = m$. 
Relation (\ref{6.4}) follows
>from restrictions ({\bf A}), ({\bf B}) and ({\bf C})
for (ee), (em) and (mm) cases respectively.
Analogously (\ref{6.5}) follows
>from ({\bf B}), ({\bf C}) and ({\bf B})
and (\ref{6.6}) follows
>from ({\bf C}), ({\bf B}) and ({\bf A})
for (ee), (em) and (mm) cases respectively. Proposition is proved.

>From relation   ({\bf B}) and $d_0 > 1$ we get
\beq{6.12}
K(a) \equiv n_{a} - 1 - N(a,a) > 0
\eeq
and hence
\beq{6.13}
(U^s,U^s) = K(a_s)  > 0.
\eeq
Thus, inequalities (\ref{5.5n}) are satisfied.

{\bf Definition 1.} The model (\ref{2.1i})
 with $\Lambda = 0$ defined on the manifold (\ref{2.10g})
 is called binary complete if for any $a_1,a_2 \in \Delta$, 
 $v_1,v_2 \in \{e,m \}$ there exists cosmological solution 
 from 5.2.1 with two intersecting p-branes defined by indices
 $s_1 = (a_1,v_1,I_1) \neq  s_2 = (a_2,v_2,I_2)$.

For binary complete model we have the maximal number
$|\Delta|(2 |\Delta| +1)$
of different types of binary configurations (formed by $|\Delta|$ electric
and $|\Delta|$ magnetic $p$-branes).

{\bf Definition 2.}  The matrix (\ref{6.8}) satisfying the
 restrictions ({\bf A}), ({\bf B}) and ({\bf C})
of Proposition 1 is called $d_0$-proper.

>From Proposition 1 and  (\ref{6.13}) we
obtain 

{\bf Proposition 2.} The  model  (\ref{2.1i}) with $\Lambda = 0$ 
defined on the manifold (\ref{2.10g}) with
$M_1 = \ldots M_n = {\bf R}$ ($n = D-1 -d_0$, $d_0 > 1$)
is binary complete if fundamental matrix  (\ref{6.8}) of this model is
$d_0$-proper.

In the next subsection we  apply 
these propositions for $D = 11$ supergravity and 
the chain of $D > 11$ models.

\subsection{\bf Example: chain of $B_D$-models}

Let us consider the action in dimension $D$
\beq{6.14}
S_D = \int_{M} d^{D}z \sqrt{|g|} \{ {R}[g] +
g^{MN} \partial_{M} \vec{\varphi} \partial_{N} \vec{\varphi}
- \sum_{a = 4}^{D-7}
\frac{1}{a!} \exp[ 2 \vec{\lambda}_{a} \vec{\varphi}] (F^a)^2 \},
\eeq
where $\vec{\varphi}  = (\varphi^1, \ldots, \varphi^l) \in {\bf R}^l$,
$\vec{\lambda}_a =(\lambda_{a1}, \ldots, \lambda_{al}) \in {\bf R}^l$,
$l = D-11$, ${\rm rank} F^a = a$, $a = 4, \ldots, D-7$.
Here vectors $\vec{\lambda}_a$ satisfy the relations
\bear{6.15}
&& \vec{\lambda}_{a} \vec{\lambda}_b = N(a,b) 
- \frac{(a - 1) (b - 1)}{D-2}, \\
\label{6.16}
&&N(a,b) = {\rm min}(a,b) - 3,
\ear
$a,b = 4, \ldots, D-7$. 

The vectors $\vec{\lambda}_a$ are linearly dependent  
\beq{6.17}
\vec{\lambda}_{D-7} = -2 \vec{\lambda}_{4}.
\eeq
For $D > 11$  vectors $\vec{\lambda}_{4}, \ldots, \vec{\lambda}_{D-8}$
are linearly independent.

The model (\ref{6.14}) contains $l$ scalar fields with negative kinetic
term (i.e. $h_{\alpha \beta} = - \delta_{\alpha \beta}$ in
(\ref{2.1i})) coupled with $l + 1$ forms. For $D = 11$ ($l= 0$) 
the model (\ref{6.14}) coincides
with truncated bosonic sector of $D = 11$ supergravity
("truncated" means without Chern-Simons term). For $D = 12$ ($l=1$)
(\ref{6.14}) coincides with truncated $D = 12$ model from \cite{KKP}
(see also \cite{IM}).
We call the model  (\ref{6.14})  $B_D$-model.

The matrix  (\ref{6.16})  is fundamental matrix (see (\ref{6.8}))
of the model (since $ \lambda_{a} \cdot \lambda_b = 
- \vec{\lambda_{a}} \vec{\lambda_b}$, $n_{a} = a$).
 
The dimensions of $p$-brane worldsheets are (see (\ref{6.7}))
\bear{6.18}
d(I) = && 3, \ldots, D -8,  \quad I \in \Omega_{a,e}, \\
\label{6.19}
       && D-5, \ldots, 6,  \quad I \in \Omega_{a,m}. 
\ear
We have $(l+1)$ electric and  $(l+1)$  magnetic $p$-branes
$p =d(I)-1$. The intersection rules are given by (\ref{6.9a}).
For $D=12$ see also \cite{IM}.

Fundamental matrix  (\ref{6.16})  is $d_0$-proper, iff $d_0 = 2$ and
hence $B_D$-model is binary complete if the manifold
decomposition (\ref{2.10g}) with
$M_1 = \ldots M_n = {\bf R}$, $n = D-3$, $d_0 =2$
is chosen. So, for spherically-symmetric solutions
the binary-completeness takes place only in "non-Tangerlini" case
 $M_0 = S^2$.

We note that in $B_D$-models 
$(U^s,U^s) = 2 > 0$, since
\beq{6.20}
K(a) =2
\eeq
(see (\ref{6.13})) for all $a =  4, \ldots, D-7$.

{\bf Remark 2.} Here two problems arise. The first one is the 
existence of the  chain of extensions
\beq{6.0a}
B_D \longleftarrow \hat{B}_D  \longleftarrow   SG_D
\longleftarrow   M_D = F_D,
\eeq
where $M_D = F_D$ is $M$- or $F$-theory in dimension $D$,
$SG_D$ is supergravitational (field) theory coinciding
with low-energy limit of $F_D$ and  $\hat{B}_D$ is 
the bosonic sector of $SG_D$. $\hat{B}_D$-action
is the sum of $B_D$-action  and  Chern-Simons terms.
The second problem is the existence 
of relations between $B_D$ models via dimensional
reduction 
\beq{6.0b}
B_{11} \longleftarrow B_{12} 
\longleftarrow  B_{13} \longleftarrow  
B_{14} \longleftarrow  \ldots .
\eeq
For $B_{11} \longleftarrow B_{12}$ see \cite{KKP}. 

We note that $B_D$ models have also rather interesting classes of Toda 
lattice solutions. The  intersection rules  for these 
"Toda  p-branes"  may be obtained using general relations 
described in the next subsection.  (To our knowledge $p$-brane solutions 
governed by open Toda lattices with $a_n$ Lie algebras were studied first 
in \cite{LPX,LMPX}.)

\subsection{\bf Generalizations to Toda lattices}

Let us consider the Lagrange system (\ref{5.31n}), (\ref{5.32n}),
where $V$ (curvature part of potential) is defined in (\ref{5.6n}).
Here we put  
\beq{6.21a}
(U^s,U^s) > 0, \quad s \in S_*.
\eeq

The Lagrangian (\ref{5.31n})  may be "block-diagonalized" in $z$-variables
(\ref{5.12n}) satisfying (\ref{5.13n}), (\ref{5.14n}), $\eta_s = +1$ and
\beq{6.21}
  B^s_{s'} z^{s'} = U^s(x),
\eeq
$s \in S_*$, instead (\ref{5.15n}).  Vectors 
$\vec{B}^s = (B^s_{s'})$ obey the relations
\beq{6.22}
\vec{B}^{s_1} \vec{B}^{s_2} = (U^{s_1},U^{s_2}), 
\eeq
$s_1, s_2 \in S_*$. In $z$-variables the Lagrangian  (\ref{5.31n})
reads
\beq{6.23}
L_Q = L_0 + L_{T,Q} + L_{f},
\eeq
where
\bear{6.24}
L_0 = - \frac{1}{2} (\dot{z}^0)^2 - \frac12w\xi_0d_0\e^{2q_0 z^0}, \\
\label{6.25}
L_{T,Q} =  \frac{1}{2} \sum_{s \in S_*} (\dot{z}^s)^2 
- 
\frac12 \sum_{s \in S_*} \eps_s Q_s^2 \exp(2 \vec{B}^{s} \vec{z}), \\
\label{6.26}
L_f =  \frac{1}{2} \eta_{ab} \dot{z}^a \dot{z}^b,
\ear
$\vec{z} = (z^s)$; $\eps_s Q_s \neq 0$, $s \in S_*$.

Thus, the problem of integrability of the Lagrange system (\ref{6.23}) is 
reduced to the problem of integrability of the 
Euclidean Toda-like system with the Lagrangian  (\ref{6.25}).

Let vectors $\vec{B}^{s}$ (or, equivalently, $U^{s}$), $s \in S_*$,
are linearly independent. According to Adler-van-Moerbeke 
criterion \cite{AM} the  equations of motion corresponding to
(\ref{6.25}) are integrable in quadratures if and only if
\beq{6.27}
\frac{2\vec{B}^{s_1} \vec{B}^{s_2}}{\vec{B}^{s_2} \vec{B}^{s_2}}
= \frac{2 (U^{s_1},U^{s_2})}{ (U^{s_2},U^{s_2})} =C_{s_1 s_2},
\eeq 
where $C = (C_{s_1 s_2})$  is the Cartan matrix for some semisimple Lie  
algebra ${\bf g}$. When ${\bf g}$ is simple $L_{T,Q}$ describes
Toda lattice corresponding to  ${\bf g}$  \cite{B}-\cite{OP}.
>From (\ref{5.4n}), (\ref{6.13}), (\ref{6.3}) and  (\ref{6.27}) 
we obtain the intersection rules
\beq{6.28}
d(I_{s_1} \cap I_{s_2}) = \Delta(s_1,s_2) + \frac{1}{2} 
K(a_{s_2})C_{s_1 s_2},
\eeq  
$s_1 \neq s_2$,
where $\Delta(s_1,s_2)$ is defined  in (\ref{6.3}) and $K(a)$ in
(\ref{6.12}).

When ${\rm rank}(\vec{B}^{s}, s \in S_*) = |S_*| - 1$ (i.e. only one
vector linearly depends upon others)  the Adler-van-Moerbeke 
criterion has the form (\ref{6.27}) with  the Cartan matrix
corresponding to some affine Lie algebra 
\cite{AM}. In this case a closed Toda lattice arises.

Thus, when relations (\ref{6.12}) and  (\ref{6.28})
are satisfied, the cosmological model is integrable since it is 
reduced effectively to  (open or closed) Toda lattice.

{\bf Uninorm models.} Let $K(a) =K $, $a \in \Delta$, and
$|S| \geq 2$. Then from  (\ref{6.28}) we get that the Cartan 
matrix is symmetric. Hence  ${\bf g}$ is of A-D-E type 
(or simply laced), i.e. it
belongs to one of $a_n$, $d_n$, $e_n$ series. In this case
$C_{s_1 s_2} = 0, -1$, for $s_1 \neq s_2$ and $-1$ takes place for some
$s_1 \neq s_2$.  Let $\Delta(s_1,s_2) \in {\bf Z}$.
Then, it follows from (\ref{6.28}) that $K$ should be even.

{\bf Example 1}. In $D = 11$ supergravity $K(a) = 2$. For 
$S = \{s_1, s_2 \}$ and ${\bf g} = a_2= sl(3)$ we get the intersection
rule
\beq{6.29}
d(I_{s_1} \cap I_{s_2}) = \Delta(s_1,s_2) -1 = 0,1,3
\eeq  
for 
$\{d(I_{s_1}), d(I_{s_2}) \}= \{3,3\}, \{3,6\}, \{6,6\}$
respectively. We see that two membranes are intersecting
in a point.  

{\bf Example 2}.  Let us consider  $D = 11$ supergravity
and put $|S| = n =3$, $d_1 = d_2 = d_3 = 3$, $d_0 = 1$.
Let $\Omega_e = \{ 1,2,3 \}$, i.e. we consider three
non-intersecting  electric
2-branes (Euclidean for $w = -1$) attached to 
$M_1, M_2, M_3$ respectively. In this case $C_{s_1 s_2} =  -1$
for $s_1 \neq s_2$  (the Dynkin diagram is a triangle) and 
the Lagrangian  $L_{T,Q}$  from (\ref{6.25}) is 
closed Toda-lattice Lagrangian corresponding to 
affine Lie algebra $ a_2^{(1)}$. (In this case 
curvature term in Lagrangian $L_0$ from (\ref{6.24}) is absent 
and  (\ref{5.14n}) should be modified.)
The corresponding solutions are expressed in terms of 
$\theta$-functions \cite{Kr}.

{\bf Example 3}. Let $K(a) = 2$ and $n_a = 1$ for all $a \in \Delta$.
>From (\ref{6.3}) and (\ref{6.28}) we get
\beq{6.30}
\chi_{s_1} \chi_{s_2} \lambda_{a_{s_1}} \cdot \lambda_{a_{s_2}}
= C_{s_1 s_2}.
\eeq
This relation implies the appearance of $E_N$ (open) Toda lattices
in maximal supergrativies in $D$ dimensions coming from
$D =11$ supergravity \cite{LMMP}.

We see, that the appearance of A-D-E algebras
seems to be rather typical for supergravitational models 
(with $K(a) =2$). 


\section{Conclusions}

In this paper we obtained exact solutions to Einstein
and  Wheeler--De Witt equations for the multidimensional cosmological 
model describing the evolution of $n$ Ricci-flat spaces
and one Einstein space $M_0$  of non-zero curvature
in the presence of composite electro-magnetic $p$-branes.
The solutions were obtained in orthogonal case (\ref{5.4n}),
when $p$-branes do not "live" in $M_0$. As an illustration
we singled out   the spherically-symmetric solutions with intersecting 
non-extremal $p$-branes.

In classical case we reduced the  model
to the Euclidean Toda-like system.  Applying the Adler-van-Moerbeke 
criterion we obtained (Toda) intersection rules  (\ref{6.28})
that imply the reduction of the  cosmological model to 
(open or closed) Toda lattice.
For uninorm models ($K(a) =K$, $a \in \Delta$) simply laced 
(A-D-E) Lie algebras appear. 
We gave some examples of reduction to Toda lattice
(e.g. closed one). Concrete exact solutions will be considered in
separate publication.

Also we developed (general) formalism of 
(orthogonal) intersections using the notions of fundamental matrix and 
binary completeness of the model and gave examples of binary complete 
models for $d_0 = {\rm dim} M_0 = 2$ ($B_D$-models). 
We may suppose that investigations of intersecting
(e.g. Toda) $p$-branes may be a powerful tool in
constructing new higher dimensional models (that may have 
supersymmetric or non-local extensions).

\begin{center}
{\bf Acknowledgments}
\end{center}

This work was supported in part
by DFG grants
436 RUS 113/7, 436 RUS 113/236/O(R) and
by the Russian Ministry for
Science and Technology,  Russian Fund for Basic Research,
project N 95-02-05785-a.

\section{Appendix}

\subsection{Appendix 1: Solutions with  Bessel functions} 

Let us consider two differential operators
\bear{a2.1}
2\hat H_0 = - \frac{\partial^2}{\partial z^2} + 2 A \e^{2qz}, \\ 
\label{a2.2}
2\hat H_1 =
- \e^{qz}\frac\partial{\partial z}
\left(\e^{-qz}\frac\partial{\partial z}\right)+ 2 A \e^{2qz}.
\ear
Equation
\beq{a2.3}
H_k \Psi_k ={\cal E} \Psi_k
\eeq 
has the following linearly independent solutions for $q \neq 0$
\bear{a2.4}
\Psi_k(z)=  \e^{kqz/2}  B_{\omega_k({\cal E})}
\left(   \sqrt{2 A} \frac{\e^{qz}}{q}\right), \\ 
\label{a2.5}
\omega_k({\cal E})=\sqrt{\frac{k}{4}- \frac{2{\cal E}}{q^2}},
\ear
where $k = 0,1$ and $B_\omega,B_\omega=I_\omega,K_\omega$
are modified Bessel function.

\subsection{Appendix 2: Classical solutions with  orthogonal vectors } 

Let
\bear{A.1}
L=\frac12<\dot x,\dot x>-\sum_{s=1}^mA_s\exp[2<b_s,x>]
\ear
be a Lagrangian, defined on $V\times V$, where $V$ is
$n$-dimensional vector space, over ${\bf R}$, $A_s\ne0$, $s=1,\dots,m$;
$m\le n$, and $<\cdot,\cdot>$ is non-degenerate real-valued quadratic
form on $V$. Let
\bear{A.2}
<b_s,b_s>\ne0 , \\  \label{A.3}
<b_s,b_l>=0, \quad s\ne l ,
\ear
$s,l=1,\dots,m$. Denote
\bear{A.4}
\eta_s=\sign<b_s,b_s> .
\ear
Then, the Euler-Lagrange equations for the Lagrangian (\ref{A.1}) have
the following solutions \cite{GIM}
\beq{A.5}
x(t)=-\sum_{s=1}^{m} \frac{b_s}{<b_s,b_s>} \ln |{f_s}(t-t_{0s})| 
+t\alpha+\beta ,
\eeq
where $\alpha,\beta\in V$,
\beq{A.6}
<\alpha,b_s>=<\beta,b_s>=0 ,
\eeq
and
\bear{A.7}
f_s(\tau)= \nn
\left|\frac{A_s}{E_s}\right|^{1/2}\sh(\sqrt{C_s}\tau),
\; C_s>0, \; A_s \eta_s<0 , \\  \label{A.8}
\left|\frac{A_s}{E_s}\right|^{1/2}\sin(\sqrt{|C_s|}\tau),
\; C_s<0, \; A_s \eta_s<0 , \\  \label{A.9}
\left|\frac{A_s}{E_s}\right|^{1/2}\ch(\sqrt{C_s}\tau),
\; C_s>0, \; A_s \eta_s>0 , \\  \label{A.10}
\left|2A_s<b_s,b_s>\right|^{1/2}\tau,
\; C_s=0, \; A_s \eta_s <0 ,
\ear
$C_s= 2 E_s<b_s,b_s>$, $E_s$, $t_{0s}$ are constants.

For the energy corresponding to the solution (\ref{A.5}) we have
\beq{A.11}
E=\sum_{s=1}^m E_s+\frac12<\alpha,\alpha> .
\eeq

For dual vectors $u^s\in V^*$ defined as 
$u^s(x)=<b_s,x>$, $\forall x \in V$, we have $<u^s,u^l>_*=<b_s,b_l>$,
where $< \cdot, \cdot>_*$ is dual form on  $V^*$.  The orthogonality 
conditions (\ref{A.6}) read 
\beq{A.14} u^s(\alpha)=u^s(\beta)=0 , 
\eeq 
$s=1,\dots,m$.

\end{document}